\documentclass[12pt]{iopart}

\usepackage{graphicx}
\usepackage{dcolumn}
\usepackage{bm}
\begin{document} 
\title{Exact bond percolation thresholds in two dimensions} 

\author{Robert M. Ziff}
\address{Michigan Center for Theoretical Physics and Department of Chemical Engineering, University of Michigan, Ann Arbor, Michigan 48109-2136, USA}
\ead{rziff@umich.edu}

\author{Christian R. Scullard}
\address{Department of Geophysical Sciences, University of Chicago, Chicago, Illinois 60637, USA}
\ead{scullard@uchicago.edu}

\date{Sept 5, 2006}

\begin{abstract}
Recent work in percolation has led to exact solutions for the site and bond critical thresholds of many new lattices. Here we show how these results can be extended to other classes of graphs, significantly increasing the number and variety of solved problems. Any graph that can be decomposed into a certain arrangement of triangles, which we call self-dual, gives a class of lattices whose percolation thresholds can be found exactly by a recently introduced triangle-triangle transformation. We use this method to generalize Wierman's solution of the bow-tie lattice to yield several new solutions. We also give another example of a self-dual arrangement of triangles that leads to a further class of solvable problems. There are certainly many more such classes.
\end{abstract}

\maketitle 

\section{Introduction}

The simple geometrical model of percolation on regular
lattices has been studied
extensively since it was first proposed nearly fifty years ago
\cite{BroadbentHammersley}. It continues to find broad
applications to such diverse problems as
understanding conductive materials \cite{Vigolo05,Grimaldi},
the fractality of coastlines \cite{Sapoval03}, networks \cite{DerenyiPallaVicsek05,CallawayNewmanStrogatzWatts00,Kalisky}, turbulence \cite{Cardy06,Bernard06}, colloids \cite{Anekal}, and the spin quantum Hall transition \cite{Gruzberg}.
Along with the Ising model of a ferromagnet and the related Potts model,
percolation serves as a paradigm of statistical mechanics.

In the usual bond percolation model each edge (bond) of an infinite graph or lattice is defined to be open with probability $p$ (and closed with probability $1-p$) independently of all other bonds. If $p$ is near $0$ most bonds are closed, but there are small and sparsely distributed open clusters. With $p$ close to $1$, the landscape is dominated by an infinite open cluster. Between these two regimes, there lies a sharply defined critical threshold, $p_c$, which is a characteristic of each lattice, at which the infinite cluster makes its first appearance. 
Finding this critical point is central to many theoretical
and computational studies of percolation. Yet, for
many years, exact thresholds were known for only a relatively
small number of two-dimensional lattices (square, triangular, honeycomb,
bow-tie...) found through duality and a technique called the 
star-triangle transformation \cite{SykesEssam,Wierman84}.

Recently, it has been recognized that the star-triangle transformation can 
be generalized for systems with correlated bonds \cite{Scullard06,ChayesLei}
or for systems composed of triangular cells more complicated
than a simple star or triangle \cite{Ziff06}.
Using this generalization, site (in the site problem we consider vertices occupied with probability $p$) and bond thresholds have been found
for many new lattices, including the ``martini" (Figure \ref{fig:martini}),
``martini-A" and ``martini-B" lattices \cite{Scullard06,Ziff06}. Recently, Wu \cite{Wu06} has shown that the transition
point of the Potts model (which includes both the Ising model and percolation
as special cases) can also be found explicitly for these three lattices -- the first new
exact transition points for the Potts model found in many years.

\begin{figure}
\begin{center}
\includegraphics[]{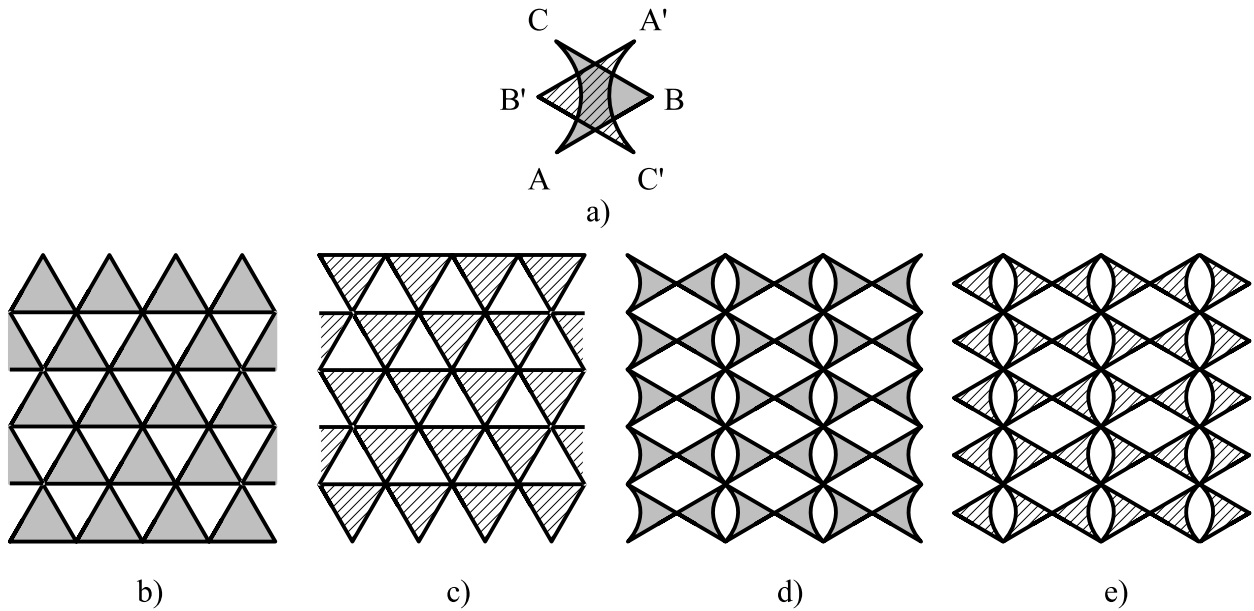}
\caption{The triangle-triangle transformation is shown in (a). The lattices in (b) and (d) are invariant under this transformation, as shown in (c) and (e).} \label{fig:triangle-triangle}
\end{center}
\end{figure}

\begin{figure}
\begin{center}
\includegraphics{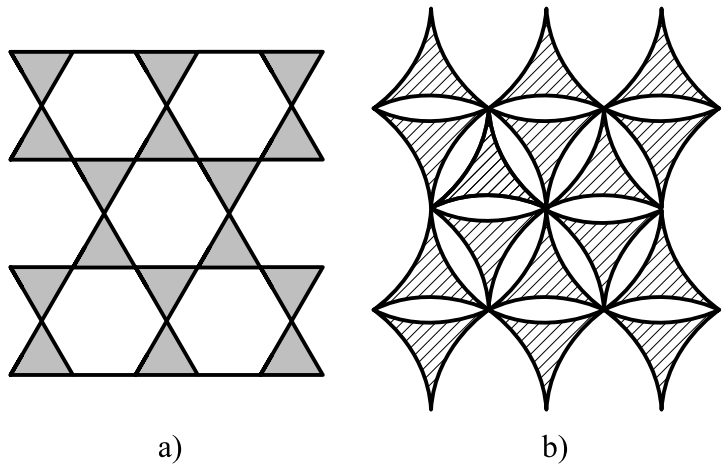}
\caption{a) The kagom\'e lattice. As clear from b) this system is not self-dual under the triangle-triangle transformation.} \label{fig:kagome}
\end{center}
\end{figure}

In this report we briefly discuss these new results, and show how they can be extended further
to systems where the triangular cells do not necessarily
lie in a triangular lattice, but in any self-dual arrangement
(in a sense defined below).
In this way we provide a derivation of the threshold
for Wierman's bow-tie lattice \cite{Wierman84} and generalizations
of it, as well as other new lattices.

\section{Triangle-triangle transformation}
We consider any system that can be decomposed into distinct 
triangles, like those shown in Fig.\ \ref{fig:triangle-triangle}. For lattices of this type that are also invariant under a certain transformation, which we call the triangle-triangle transformation, it is possible to determine exactly the bond threshold. The triangle-triangle transformation is illustrated in Figure\ \ref{fig:triangle-triangle}(a) where we superpose a reversed (hatched) triangle over each (shaded) triangle of the original lattice (Fig.\ \ref{fig:triangle-triangle}(c)). If the lattice formed by the hatched triangles is the same as the original, we say that the lattice is self-dual under the triangle-triangle transformation. The shaded triangle (the generator) can represent any network of bonds, correlated bonds, and sites, as long as they are contained between three vertices. This means that every self-dual arrangement of triangles gives an infinite family of solvable lattices. Figure\ \ref{fig:kagome} shows an example, the kagom\'e arrangement, that can be decomposed into triangles but is not self-dual under the triangle-triangle transformation. Although some special lattices fall into this arrangement, such as the double bond honeycomb lattice, and can be solved exactly, this is not true for every choice of generator. For example, choosing a simple triangle with uncorrelated bonds in Figure\ \ref{fig:kagome}(a), gives the kagom\'e lattice. This lattice's bond threshold cannot be found by the following method and, in fact, the problem remains unsolved \cite{ScullardZiff06}.

\begin{figure}
\begin{center}
\includegraphics{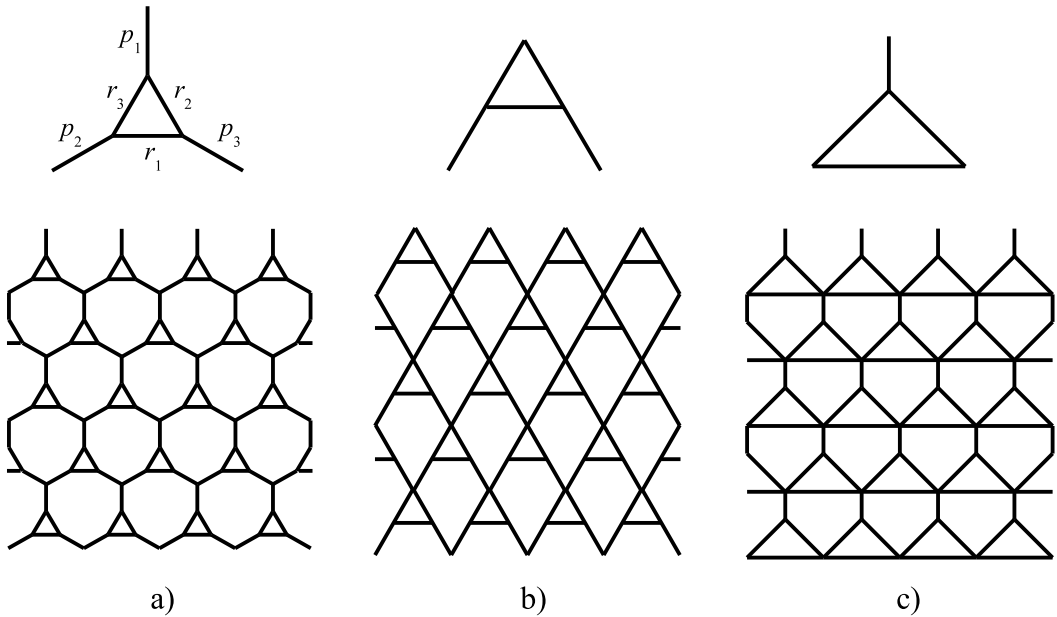}
\caption{Some examples of lattices created by using different generators in the arrangement of Fig. \ref{fig:triangle-triangle}(b). a) The martini lattice, b) The martini-A lattice, c) the martini-B lattice.} \label{fig:martini}
\end{center}
\end{figure}

To find the general condition for criticality, consider any self-dual arrangement with
vertices $A,B,C$ on the original triangles, and $A', B', C'$ on the dual
triangles as in Fig.\ \ref{fig:triangle-triangle}.  To be
at the critical threshold, we require two things (in exact analogy with the
usual star-triangle transformation): (1) that the connection probabilities
between the vertices of $A,B,C$ and those of the vertices of $A',B',C'$
are the same, and (2) that  $A,B,C$ and $A',B',C'$ are dual to each
other.  The former implies that the overall connection probabilities
between points on the lattice and corresponding ones on
the (identical) dual lattice are
the same, and the latter implies that this equi-probability point
must be the critical point (because only at the critical point
can a lattice and its dual have identical correlations).

To state these conditions mathematically,
we let  $P(A,B,C)$ equal the probability that $A$, $B$, and $C$ on a triangle
are connected, $P(A, B, \overline C)$ equal the probability that $A$
and $B$ are connected but not $C$, etc., and $P(\overline A, \overline B, \overline C)$
equal the probability that none are connected. Likewise define
$P'(A', B', C')$ etc. for the
connection probabilities on the dual triangles.  Then the condition
of equality of connection probabilities is expressed by
\begin{eqnarray}
P(A,B,C) &=& P'(A',B',C') \label{eq:eqcon1}\\
P(A,B,\overline C) &=& P'(A',B',\overline C') \\
P(\overline A, \overline B, \overline C) &=& P'(\overline A', \overline B', \overline C') \label{eq:eqcon3}
\end{eqnarray}
Secondly, we require that the two triangles $A,B,C$ and $A',B',C'$
are dual to each other.  This requires
\begin{eqnarray}
P(A,B,C) &=& P'(\overline A',\overline B',\overline C') \label{eq:dual1}\\
P(A,B,\overline C) &=& P'(A',B',\overline C') \\
P(\overline A, \overline B, \overline C) &=& P'(A', B', C') \label{eq:dual3}
\end{eqnarray}
For example, the first says simply that if $A, B,$ and $C$ are connected
to each other, then $A'$, $B'$, and $C'$ must all be isolated. This is because the dual triangular unit $A',B',C'$ is created by
drawing bonds intersecting all the bonds of $A, B, C$, and placed
down with complementary probilities $P_i = 1 - p_i$. Thus, whenever there is an occupied bond on the original lattice, the dual-lattice bond is not occupied, and vice-versa. Therefore, if $A$, $B$, and $C$
are all connected, the bonds between $A'$, $B'$ and $C'$
must necessarily be all unconnected.  The same kind of argument
holds for the other conditions. 

Comparing the two sets of equations (\ref{eq:eqcon1})-(\ref{eq:eqcon3}) and (\ref{eq:dual1})-(\ref{eq:dual3}), we see that
the middle (two-point) equations are identical.  The first and third
equations yield the requirement that the probability that all
the vertices connect is the same as the probability that none of them connect:
\begin{equation}
P(A,B,C) = P(\overline A, \overline B, \overline C). \label{criterion}
\end{equation}
This simple general relation determines the critical point, and
was given independently in Refs. [15] and [16].  It is only
a function of the arrangement of bonds in the individual
triangles and not dependent upon how the triangles are arranged
to form a lattice, as long as they are arranged in an overall
self-dual way as those in Fig. 1.
In our previous work \cite{Ziff06},
(\ref{criterion}) was applied only to systems of triangular cells that were on a triangular
array (Fig.\ \ref{fig:triangle-triangle}(b)). The lattices we studied, along with their generators, are shown in Figure \ref{fig:martini}. As already mentioned, the site thresholds for these lattices can also be found exactly. In fact, they were initially derived by a different method \cite{Scullard06} but were then shown to fall into the current framework by using generators that involve sites as well as bonds \cite{Ziff06}. These bond and site thresholds are given in Tables \ref{table:oldbondthresholds} and \ref{table:oldsitethresholds}. Here we apply (\ref{criterion}) to any self-dual set of triangles such as those in Figs.\ \ref{fig:triangle-triangle}(d) and \ref{fig:secondselfdual}.
The generalization of (\ref{criterion}) to all self-dual arrangements 
is immediately obvious from the arguments of Refs.\ \cite{Ziff06,ChayesLei}, and was suggested by the earlier work of Wierman on the bow-tie lattice \cite{Wierman84}.

\section{Martini and related lattices}
The result from which all our exact solutions will follow is the critical threshold of the martini lattice in which each bond of the generator is assigned a different probability, as shown in Figure\ \ref{fig:martini}(a). This results in a critical surface rather than a critical point and application of (\ref{criterion}) implies that it is given by \cite{Wu06,ScullardZiff06}
\begin{eqnarray}
1 &-& p_1 p_2 r_3 - p_2 p_3 r_1 - p_1 p_3 r_2 - p_1 p_2 r_1 r_2 \nonumber \\
&-& p_1 p_3 r_1 r_3 - p_2 p_3 r_2 r_3 + p_1 p_2 p_3 r_1 r_2 \nonumber \\
&+& p_1 p_2 p_3 r_1 r_3 + p_1 p_2 p_3 r_2 r_3 + p_1 p_2 r_1 r_2 r_3 \nonumber \\
&+& p_1 p_3 r_1 r_2 r_3 + p_2 p_3 r_1 r_2 r_3 - 2 p_1 p_2 p_3 r_1 r_2 r_3 = 0  \ .
\label{eq:martini}
\end{eqnarray}
If we set all bonds equal to $p$, we get the condition
$
(2 p^2-1)(p^4-3 p^3+2 p^2 +1)=0
$,
which has solution on $[0,1]$ $p_c=1/\sqrt{2}$. Setting $p_1=1$ and $p_2=p_3=r_1=r_2=r_3=p$ gives the martini-A lattice, and setting $p_2=p_3=1$ and $p_1 = r_1 = r_2 = r_3 = p$ gives the martini-B lattice, as listed in Table\ \ref{table:oldbondthresholds}.
\begin{figure}
\begin{center}
\includegraphics[]{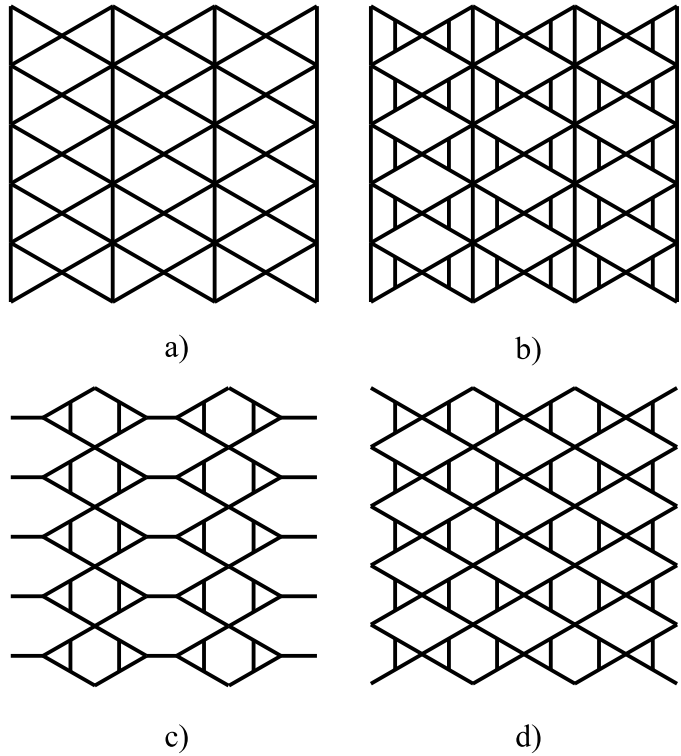}
\caption{Some examples of generalized bow-tie lattices.} \label{fig:bowtie_lattices}
\end{center}
\end{figure}
\begin{figure}
\begin{center}
\includegraphics[]{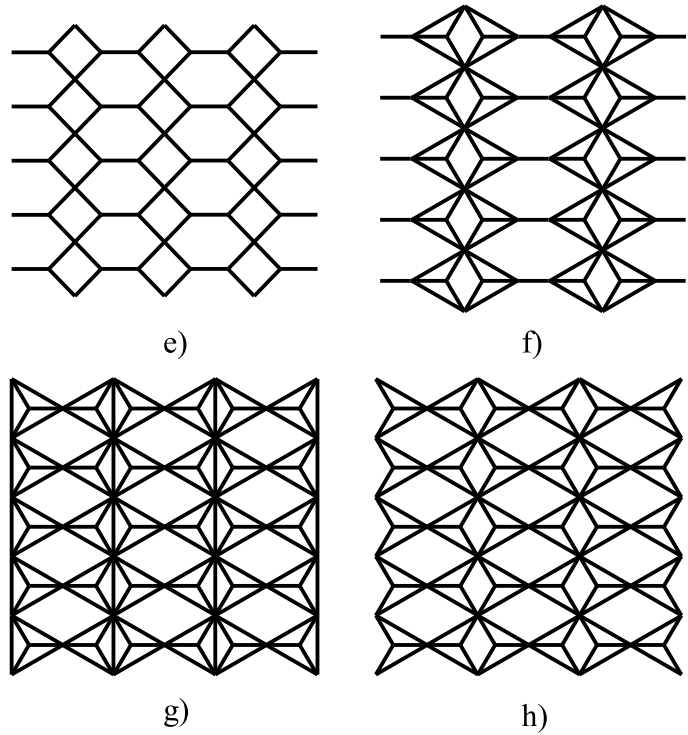}
\caption{The duals of the lattices shown in Fig.\ \ref{fig:bowtie_lattices}.} \label{fig:bowtie_duals}
\end{center}
\end{figure}

\begin{table}
\begin{center}
\begin{tabular}{lll}
{lattice} & {\bf $p_c^{\mathrm{bond}}$} & equation\\
\hline
\\
martini & $1/\sqrt{2} \approx 0.707107...$  & $(2 p^2-1)(p^4-3 p^3+2 p^2 +1)=0$\\
martini-A& $0.625457...$ &  $p^5-4 p^4+3 p^3 + 2 p^2 - 1 = 0$\\
martini-B& $1/2$ &$(2 p -1)(p^2-p-1)=0$\\
\end{tabular}
\end{center}
\caption{Bond percolation thresholds of the lattices in Fig.\ \ref{fig:martini} and their associated polynomials derived from equation (\ref{eq:martini}).}
\label{table:oldbondthresholds}
\end{table}

\begin{table}
\begin{center}
\begin{tabular}{lll}
{lattice} & {\bf $p_c^{\mathrm{site}}$} & equation\\
\hline
\\
martini & $0.764826...$ & $p^4-3 p^3 +1=0$\\
martini-A& $1/\sqrt{2} \approx 0.707107...$ & $2 p^2-1 = 0$\\
martini-B& $1/\phi \approx 0.618034...$ & $p^2+p-1=0$\\
\end{tabular}
\end{center}
\caption{Site percolation thresholds of the lattices in Fig.\ \ref{fig:martini} and their associated polynomials. $\phi=(1 + \sqrt{5})/2 = 1.618034...$ is the golden ratio, which makes an interesting appearance here.}
\label{table:oldsitethresholds}
\end{table}
Taking $r_i = 1$, we get the 
result for the critical surface of the general honeycomb lattice \cite{SykesEssam},
$
1 - p_1 p_2 - p_1 p_3 - p_2 p_3 + p_1 p_2 p_3 =0 \ ,
$
and taking $p_i = 1$ yields the critical
surface of the general triangular lattice \cite{SykesEssam},
$
1 - r_1 - r_2 - r_3 + r_1 r_2 r_3 =0 \ .
$
Setting all probabilities equal in the last two expressions gives polynomial equations with the well-known solutions $p_c(\mathrm{honeycomb})=1-2 \sin \pi/18$ and $p_c(\mathrm{triangular})=2 \sin \pi/18$. It is a general fact that thresholds found via (\ref{criterion}) are solutions of polynomials with integer coefficients.

\section{Bow-tie class of lattices}
Although equation (\ref{eq:martini}) was originally derived for the martini lattice, it gives the critical surface of any lattice in which the martini generator is arranged in a self-dual way. Here we apply the result to the arrangement given in Fig.\ \ref{fig:triangle-triangle}(d). Taking $p_i=1$ in this case gives the lattice of Fig.\ \ref{fig:triangle-triangle}(d) where the shaded triangle is just a simple triangular cell. This lattice has a double bond where the bases of abutting triangles join. However, we can use a trick, due to Wierman \cite{Wierman84}, to combine these two bonds by setting the probability of each to be $r_3 = 1 - \sqrt{1 - p}$.
The probability of traversing the double bonds is now just $p$, which effectively combines them into one. With this replacement, we get Wierman's bow-tie lattice (Fig.\ \ref{fig:bowtie_lattices}(a)). Setting the single bonds, $r_1$ and $r_2$, to $p$ gives the condition
$
1-p-6 p^2+6 p^3-p^5=0
$,
with solution in $[0,1]$ $p_c = 0.404518...$ as found by Wierman. But we can just
as well make the basic triangular cell the ``A" generator (Fig.\ \ref{fig:martini}(b)) which produces
the lattice shown in Fig.\ \ref{fig:bowtie_lattices}(d), with a transition point identical to that of the regular ``A" lattice (given in Table \ref{table:oldbondthresholds}) since no double bond arises in this case. Of course, there is an infinite variety of additional 
lattices that can be solved by this procedure and it becomes a somewhat subjective matter to decide which ones are most interesting; some of our choices are shown in Fig.\ \ref{fig:bowtie_lattices}, with their duals in Fig.\ \ref{fig:bowtie_duals} and thresholds in Table \ref{table:thresholds}. For (c), we use a martini generator with $p_1 = \sqrt{p}$
and $p_2 = p_3 = r_1 = r_2 = r_3 = p$, while for (b) we use an ``A" generator with 
an extra bond on the bottom,
which results in a doubled bond that must be combined.

\begin{figure}
\begin{center}
\includegraphics[]{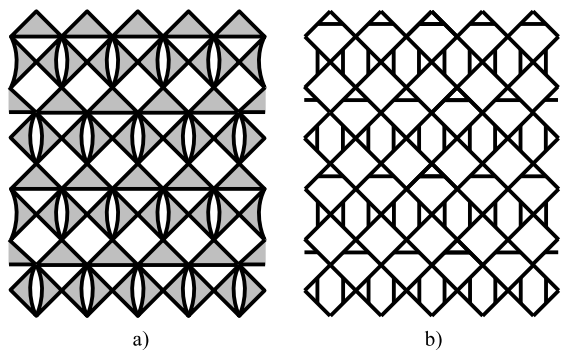}
\caption{a) Another self-dual triangular lattice under the triangle-triangle transformation. b) a particular example of a) using the ``A'' lattice generator} \label{fig:secondselfdual}
\end{center}
\end{figure}
\begin{table}
\begin{center}
\begin{tabular}{lll}
{lattice} & {\bf $p_c^{\mathrm{bond}}$} & equation\\
\hline
\\
a)& $0.404518$  & $1-p-6 p^2+6 p^3-p^5=0$  \cite{Wierman84}\\
b)& $0.533213$ &  $1 - p - 2p^3 - 4p^4 - 4 p^5 + 15 p^6$\\
&&$+ 13 p^7 - 36 p^8 + 19 p^9 + p^{10} + p^{11}=0$ \\
c)& $0.672929$ &$1 - 2 p^3 - 2 p^4 - 2 p^5 - 7 p^6 + 18 p^7$\\
&&$ + 11 p^8 - 35 p^9 + 21 p^{10} - 4 p^{11}=0$ \\
d)& $0.625457$ & $1 - 2 p^2 - 3 p^3 + 4 p^4 - p^5=0$\\
e)& $0.595482$ &  $= 1 - p_c(a)$\\
f)& $0.466787$ & $= 1 - p_c(b)$\\
g)& $0.327071$ & $= 1 - p_c(c)$\\
h)& $0.374543$ & $= 1 - p_c(d)$\\
\end{tabular}
\end{center}
\caption{Thresholds of the bow-tie lattices in Figs.\ \ref{fig:bowtie_lattices} and \ref{fig:bowtie_duals}}
\label{table:thresholds}
\end{table}

In Fig.\ \ref{fig:secondselfdual}(a) we show another arrangement of triangles that is self-dual under the triangle-triangle
transformation, defining another family of solvable lattices. For example, replacing
the shaded triangles by the ``A" generator, one can get an interesting variation 
on the ``A" lattice with the same threshold, shown in Fig.\ \ref{fig:secondselfdual}(b). Many similar self-dual arrangements can be constructed.

Thus, we have shown that a great variety of lattices
with exactly determined percolation thresholds can now be
constructed.  We also see that there are many variations of a
given lattice (such as the lattices with the A-lattice threshold
shown in Figs. 3(b), 4(d), and 6(b)) that share the same threshold,
even though the bonds are in different arrangements.
Even after decades of study, the percolation model continues to yield new
exact results.

\section*{Acknowledgments}

R.Z. acknowledges support from NSF Grant No. DMS-0553487.

\section*{References}

\bibliography{scullard_ziffv4}

\begin{thebibliography}{10}

\bibitem{BroadbentHammersley}
S.~R. Broadbent and J.~M. Hammersley.
\newblock {\em Proc. Cambridge Phil. Soc.}, 53:629--641, 1957.

\bibitem{Vigolo05}
B.~Vigolo, C.~Coulon, M.~Maugey, C.~Zakri, and P.~Poulin.
\newblock {\em Science}, 309(5736):920--923, 2005.

\bibitem{Grimaldi}
C.~Grimaldi and I.~Balberg.
\newblock {\em Phys. Rev. Lett.}, 96:066602, 2006.

\bibitem{Sapoval03}
B.~Sapoval, A.~Baldassarri, and A.~Gabrielli.
\newblock {\em Phys. Rev. Lett.}, 93:098501, 2003.

\bibitem{DerenyiPallaVicsek05}
I.~Der\'enyi, G.~Palla, and T.~Vicsek.
\newblock {\em Phys. Rev. Lett.}, 94:160202, 2005.

\bibitem{CallawayNewmanStrogatzWatts00}
D.~S. Callaway, M.~E.~J. Newman, S.~H. Strogatz, and D.~J. Watts.
\newblock {\em Phys. Rev. Lett.}, 85:5468, 2000.

\bibitem{Kalisky}
T.~Kalisky and R.~Cohen.
\newblock {\em Phys. Rev. E}, 73:035101(R), 2006.

\bibitem{Cardy06}
J.~L. Cardy.
\newblock {\em Nature Physics}, 2:67--68, 2006.

\bibitem{Bernard06}
D.~Bernard, G.~Boffetta, A.~Celani, and G.~Falkovich.
\newblock {\em Nature Physics}, 2:124--128, 2006.

\bibitem{Anekal}
S.~G. Anekal, P.~Bahukudumbi, and M.~A. Bevan.
\newblock {\em Phys. Rev. E}, 73:020403(R), 2006.

\bibitem{Gruzberg}
I.~A. Gruzberg, A.~W.~W. Ludwig, and N.~Read.
\newblock {\em Phys. Rev. Lett.}, 82(22):4524--4527, 1999.

\bibitem{SykesEssam}
M.~F. Sykes and J.~W. Essam.
\newblock {\em J. Math. Phys.}, 5(8):1117, 1964.

\bibitem{Wierman84}
J.~C. Wierman.
\newblock {\em J. Phys. A}, 17:1525, 1984.

\bibitem{Scullard06}
C.~R. Scullard.
\newblock {\em Phys. Rev. E}, 73(1):016107, 2006.

\bibitem{ChayesLei}
L.~Chayes and H.~K. Lei.
\newblock {\em J. Stat. Phys.}, 122(4):647--670, 2006.

\bibitem{Ziff06}
R.~M. Ziff.
\newblock {\em Phys. Rev. E}, 73:016134, 2006.

\bibitem{Wu06}
F.~Y. Wu.
\newblock {\em Phys. Rev. Lett.}, 96:090602, 2006.

\bibitem{ScullardZiff06}
C.~R. Scullard and R.~M. Ziff.
\newblock {\em Phys. Rev. E}, 73:045102(R), 2006.

\end{thebibliography}

\end{document}